\documentclass[prc,nofootinbib,twocolumn]{revtex4}
\usepackage{amssymb,amsmath}
\usepackage{appendix}
\usepackage{graphicx}
\newcommand{\be}{\begin{equation}}
\newcommand{\ee}{\end{equation}}
\newcommand{\bea}{\begin{eqnarray}}
\newcommand{\eea}{\end{eqnarray}}

\def\Xint#1{\mathchoice
   {\XXint\displaystyle\textstyle{#1}}%
   {\XXint\textstyle\scriptstyle{#1}}%
   {\XXint\scriptstyle\scriptscriptstyle{#1}}%
   {\XXint\scriptscriptstyle\scriptscriptstyle{#1}}%
   \!\int}
\def\XXint#1#2#3{{\setbox0=\hbox{$#1{#2#3}{\int}$}
     \vcenter{\hbox{$#2#3$}}\kern-.5\wd0}}

\def\dashint{\Xint-}

\begin{document}

\title{Effective field theory as a limit of $R$-matrix theory for light
nuclear reactions}

\author{Gerald M.\ Hale}
\author{Lowell S.\ Brown}
\author{Mark W.\ Paris}
\email{mparis@lanl.gov}
\affiliation{Los Alamos National Laboratory, 
MS B283, Los Alamos, New Mexico 87545}
\date{\today}

\begin{abstract}
We study the zero channel radius limit of Wigner's $R$-matrix theory
for two cases, and show that it corresponds to non-relativistic
effective quantum field theory. We begin with the simple problem of
single-channel $np$ elastic scattering in the $^{1}\! S_0$ channel.
The dependence of the $R$ matrix width $g^2$ and level energy
$E_\lambda$ on the channel radius $a$ for fixed scattering length
$a_0$ and effective range $r_0$ is determined. It is shown that these
quantities have a simple pole for a critical value of the channel
radius, $a_p=a_p(a_0,r_0)$.  The $^3$H$(d,n)^4$He reaction cross
section, analyzed with a two-channel effective field theory in the
previous paper, is then examined using a two-channel, single-level
$R$-matrix parameterization. The resulting $S$ matrix is shown to be
identical in these two representations in the limit that $R$-matrix
channel radii are taken to zero.  This equivalence is established by
giving the relationship between the low-energy constants of the
effective field theory (couplings $g_c$ and mass $m_*$) and the
$R$-matrix parameters (reduced width amplitudes $\gamma_c$ and level
energy $E_\lambda$). An excellent three-parameter fit to the observed
astrophysical factor $\overline{S}$ is found for `unphysical' values
of the reduced widths, $\gamma_c^2<0$.
\end{abstract}

\maketitle

\section{Introduction}
\label{sec:intro}
In the previous companion paper \cite{BH13}, a two-channel effective
quantum field theory (EFT) expression for the $^3$H$(d,n)^4$He reaction 
cross
section at low energies was derived. It gives a very good description
of the experimental data over the resonance region using only three
parameters. In this paper, we examine its relation to the $R$-matrix
theory, in the limit where the channel radii are taken to vanish.

We first consider the simpler problem of elastic $np$ scattering in
the $^{1}\!S_0$ channel; it is free of any bound state and can
be treated with a single channel $R$ matrix. This preliminary $np$
study provides insight into the channel radius dependence of the
single-channel $R$ matrix parameters.

The main focus of this study is a two-channel, single-level $R$-matrix
description of the $^3$H$(d,n)^4$He cross section, which is dominated
by an $^5$He resonant contribution.  We consider such a description
both for the case of finite channel radii and in the limit as the
channel radii approach zero. We find, as have others \cite{Resler92,
Karnakov90,Karnakov91}, that an excellent fit to the experimental
reaction data can be obtained with a single $R$-matrix level using
channel radii that are on the order of the range of nuclear forces.
This is consistent with the usual interpretation that the channel
radii represent, in a qualitative sense, the short but non-zero range
over which these forces act. It may then be somewhat surprising,
however, that the high quality of the fit to the data is maintained
with much smaller radii. Indeed, the excellent agreement persists for
channel radii of {\em zero} if the square of the reduced widths are
allowed to become (unphysically) negative.  Remarkably, in this limit
the $R$-matrix expression reduces identically to that determined in
the EFT approach.

\begin{figure*}
\includegraphics[width=5in,clip]{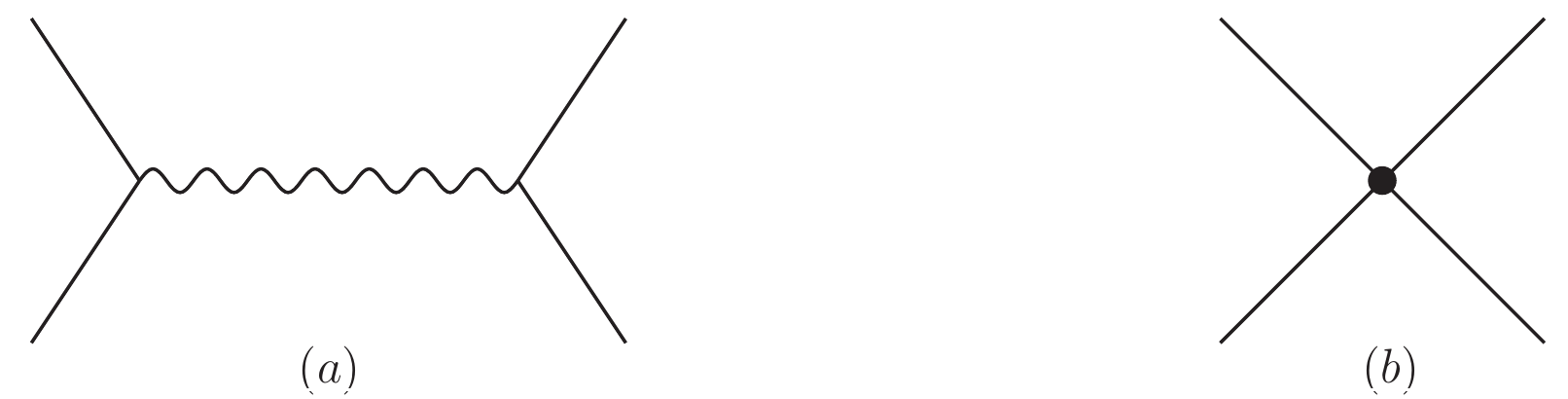} 
\caption{\label{fig:lvc} Local (a) and contact (b) interaction
mechanisms in the effective field theory approach. The equivalence of
the zero channel radius $R$-matrix description, discussed in the text,
to the EFT of Ref.\cite{BH13} demonstrates that it may be applied to
theories with
local interactions. Note that these statements should not be confused
with those relating to theories with a contact interaction, as in (b).}
\end{figure*}

Given that the channel radii are associated with the range of the
nuclear force, this identity establishes that the $R$-matrix
description can be applied to theories with {\em local} couplings, as
in Fig.~\ref{fig:lvc}a (to be contrasted with the {\em contact}
coupling of Fig.~\ref{fig:lvc}b).  The channel radii dependence of the
$R$-matrix parameters, as the size of the interior region vanishes,
gives insight into how the Lagrangian of the EFT description becomes
``unphysical."

The low-energy $np$ scattering is treated in Section \ref{sec:np} as a
simple example, which may be expressed analytically, to obtain insight
into the channel-radius dependence of the $R$-matrix parameters.  In
this case, the physical values of the $^{1}\! S_0$ scattering length
and effective range are given by real $R$-matrix parameters at radii
down to a critical channel radius of $a_p \simeq 1.30$ fm, but below
that radius (and continuing to zero radius) the reduced-width
amplitude $\gamma$ becomes pure imaginary, with $\gamma^2 < 0$. 

In Section \ref{sec:dt} we discuss the main result of the current
study, a two-channel, single-level formula for the total
$^3$H$(d,n)^4$He reaction cross section. We describe, in Subsection
\ref{subsec:fcr}, its application to the description of the
reaction at finite channel radii with a four parameter
fit. The derivation of a dispersion relation for the logarithmic
derivative of the external outgoing-wave solution that is relevant to
this development and, to our knowledge previously unpublished, is
given in the Appendix. We then describe the dependence of the
parameters of the fit on variations in the channel radii. In
Subsection \ref{subsec:zrl} we show how the zero-radius limit of the
$R$-matrix expression reduces to the one given by EFT in the companion
paper.

Finally, we discuss the implications of our work and give conclusions
in Section IV.  The notation has been changed somewhat from the
previous paper to correspond with that used more commonly in the
$R$-matrix literature for nuclear reactions.  For example, the channel
labels will be shortened so that $d$ means $dt$ (or $d+^3$H) 
and $n$ stands for
$n\alpha$ (or $n+^4$He). 
Other notational differences will be indicated as needed.

\section{Single-level Effective Range Expansion for $np$ Scattering}
\label{sec:np}

We investigate the dependence of the $R$-matrix parameters on the
channel radius $a$, in limit that $a\to 0$ for a simple example:
single-channel scattering in a neutral $S$-wave such as neutron-proton
scattering.
   
We assume that the low-energy scattering is described by a  
single-level $R$-matrix
\be
R(E;a)=\frac{\gamma^2(a)}{E_\lambda(a)-E} \,,
\ee
for which the scattering matrix is
\be
e^{2i\delta_0(E)} = e^{-2ika} \frac{1 + i ka \, R(E;a)}{1 - i ka \, R(E;a)} \,,
\label{S}
\ee
where $k=\sqrt{2\mu E}/\hbar$ is the wave number in the
center-of-mass, $\mu$
the reduced mass of the scattering pair, and $E_\lambda$ the energy
eigenvalue of the level. Instead of the reduced width amplitude
$\gamma$, we shall use $g^2 = a \, \gamma^2$ because $g^2$ remains
finite in the zero channel radius limit\footnote{See the discussion in
   Subsection \ref{subsec:zrl} and the footnote \ref{foot}. 
   In this connection, it is interesting to note
   that this is the convention with which Wigner and
   Eisenbud \cite{WE47} originally defined reduced widths in the $R$
   matrix.  Also notable is the fact that, in a conversation with one
   of the authors (G.M.H.) in 1975, Prof.\ Wigner mentioned that he
   was thinking about what $R$-matrix theory looks like at zero
   radius. It is likely that he was pondering at that time the sort of
extension to local interactions that we consider here.},
$a \to 0$.

If we simply take the channel radius to vanish, $a \to 0$, then the
transition amplitude becomes
\be
i (e^{2i\delta_0(E)} - 1)
= 2 \frac{g^2 k}{E-E_\lambda + i g^2 k} \,.
\label{amp}
\ee
This is proportional to the transition amplitude in the
non-relativistic effective quantum field theory for two particles
interacting via a single scalar intermediate field with
energy $E_\lambda$.  The zero of the energy scale, $E=0$, corresponds
to vanishing relative motion of the scattering particles. In this
field theory context, $g$ is, up to a conventional overall factor, the
coupling constant of the scattering particles interacting with the
intermediate field. The scattering amplitude \eqref{amp} can be rewritten
\be
k \cot\delta_0(E) = \frac{1}{g^2} \left(E_\lambda - E \right) \,.
\ee
This is precisely the effective range expansion
\be
k \cot\delta_0(E)=-\frac{1}{a_0}+\frac{1}{2}r_0 k^2 \,,
\ee
with the identifications
\be
\label{eqn:a0}
a_0 = - \frac{g^2}{E_\lambda}
\ee
for the scattering length, and
\be
\label{eqn:r0}
r_0 = - \frac{\hbar^2}{g^2 \mu}
\ee
for the effective range. The two-term effective range approximation 
for $k\cot\delta_0(E)$ is exact in this case.

We may, in general, have a low-energy cross section that results in
either a positive or negative scattering length.
Equations \eqref{eqn:a0} and \eqref{eqn:r0} demonstrate that, in the
zero channel-radius limit, $a\to 0$, the sign of the $a_0$ is 
determined by whether the $R$-matrix parameters,
$E_\lambda$ and $g^2$, of like ($a_0<0$) or differing ($a_0>0$) sign. 

Turning to the effective range, $r_0$ we see that
for a real coupling strength $g$, the effective range parameter
is necessarily negative, $r_0<0$, independent of the sign of
$E_\lambda$. A positive effective range, which is the predominant
situation -- as in $^{1}\!S_0$ neutron-proton scattering -- may be
obtained by the formal device of taking the coupling strength to be
purely imaginary, $g \to ig$. Although this results in a structure
that is unphysical from both the field-theoretic and $R$-matrix
perspectives, it is an acceptable procedure for our limited objective
of obtaining a low-energy description. In particular, the transition
amplitude, Eq.\eqref{amp} continues to satisfy unitarity under this
transformation, which is equivalent to $ g^2 \to - g^2$. It is clear
from Eq.~\eqref{amp} that the change $g^2 \to - g^2$ is also 
equivalent to $\left( E - E_\lambda \right) \to \left(
E_\lambda - E \right)$. In the field theory description, this is a
transformation that changes the sign of the intermediate field's
unperturbed (free-field) propagator which is brought about by changing
the sign of the intermediate field's free Lagrangian\footnote{In field
   theory language, the change $g \to ig$ is equivalent to a field
   redefinition of the independent intermediate field operators, 
   $\psi \to -i \psi, \psi^\dag \to -i\psi^\dag$.  This redefinition 
   changes the sign of the free-field Lagrangian for $\psi$ since this 
   term is proportional to $\psi^\dag\cdots\psi$.  While the field
   redefinition removes the appearance of a non-Hermitian interaction
   Lagrangian, the free-field Lagrangian is not consistent with the
   positivity postulates of quantum field theory.}.  It is
conventional, in work that applies effective quantum field theory to
nuclear physics problems \cite{EHM09}, to employ the convention that
the sign of the free-field intermediate Lagrangian is used with a
``wrong sign'' to obtain a positive effective range parameter, and so
this is the convention used in our preceding paper \cite{BH13}.
However, in our work here, it proves convenient to use the equivalent
method of using an imaginary coupling constant.

We now wish to examine the $a$ dependence in detail and, in
particular, the character of the $a \to 0$ limit.  We return to the
general $R$-matrix expression \eqref{S} with $a\ne 0$, which may 
be written as
\be
k \cot\delta_0(E) 
= \frac{E_\lambda - \frac{\hbar^2 k^2}{2 \mu} + k g^2 \tan ka}
{g^2 - (E_\lambda - \frac{\hbar^2 k^2}{2 \mu}) \frac{1}{k} \tan ka}.
\ee
This gives the scattering length
\be
 a_0 = a - \frac{g^2}{E_\lambda} \,,
\label{length}
\ee 
and the effective range
\be
r_0 = 
\frac{ 2 a^3 E_\lambda^2 / 3 -2 a^2 E_\lambda g^2 + 2 a g^4 - g^2 
\hbar^2/\mu}
           {(g^2 - a E_\lambda )^2} \,.
\label{range}
\ee

We study $g^2$ and $E_\lambda$ as functions of $a$ for fixed
values of $a_0$ and $r_0$. To this end, we use the condition
$g^2/E_\lambda=a-a_0$ from Eq.~\eqref{length} in Eq.~\eqref{range} to
obtain
\begin{align}
\frac{1}{2}r_0 a_0^2
=\frac{1}{3} a^3+(a_0-a)\left(aa_0+\frac{\hbar^2}{2\mu E_\lambda}\right) \,,
\end{align}
which gives
\be
E_\lambda(a)=\frac{\hbar^2(a_0-a)}{2\mu[r_0a_0^2/2-a^3/3-aa_0(a_0-a)]} 
\,,
\label{Elambda}
\ee
and
\begin{align}
g^2(a) &=(a-a_0)E_\lambda(a) \nonumber \\
       &=-\frac{\hbar^2(a_0-a)^2}{2\mu[r_0a_0^2/2-a^3/3-aa_0(a_0-a)]} \,.
\label{g2}
\end{align}
The denominator of Eqs.~\eqref{Elambda} and \eqref{g2} is
a cubic polynomial in $a$ with real coefficients. Hence it must have
at least one real pole in $a$. In fact, there is a single real pole at $a_p$ 
that is given by
\begin{align}
   \label{eqn:ap}
   a_p &= a_0 +
   \left\{a_0^3\left[\frac{3r_0}{2a_0}-1\right]\right\}^{1/3}.
\end{align}

\begin{figure}
\begin{center}
\includegraphics[width=3.4in, trim=50 50 50 50]{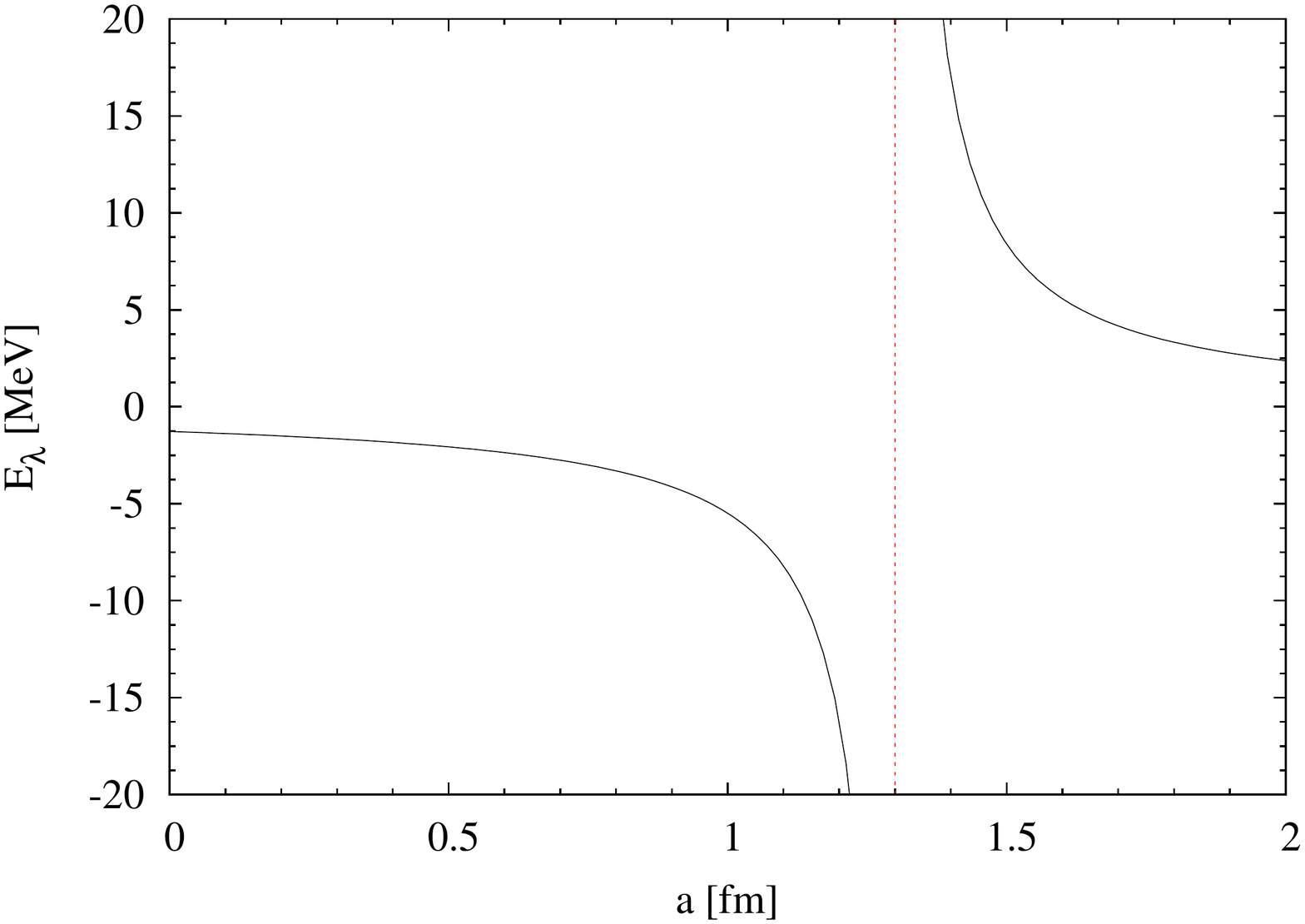} 
\caption{\label{fig:npelvsa}
Behavior of $E_\lambda(a)$ as a function of the channel radius $a$ when 
the values of the singlet $np$ scattering length and effective range 
are kept fixed. The vertical dotted line denotes the position of
the pole at $a_p=1.30$ fm.}
\includegraphics[width=3.4in, trim=50 50 50 50]{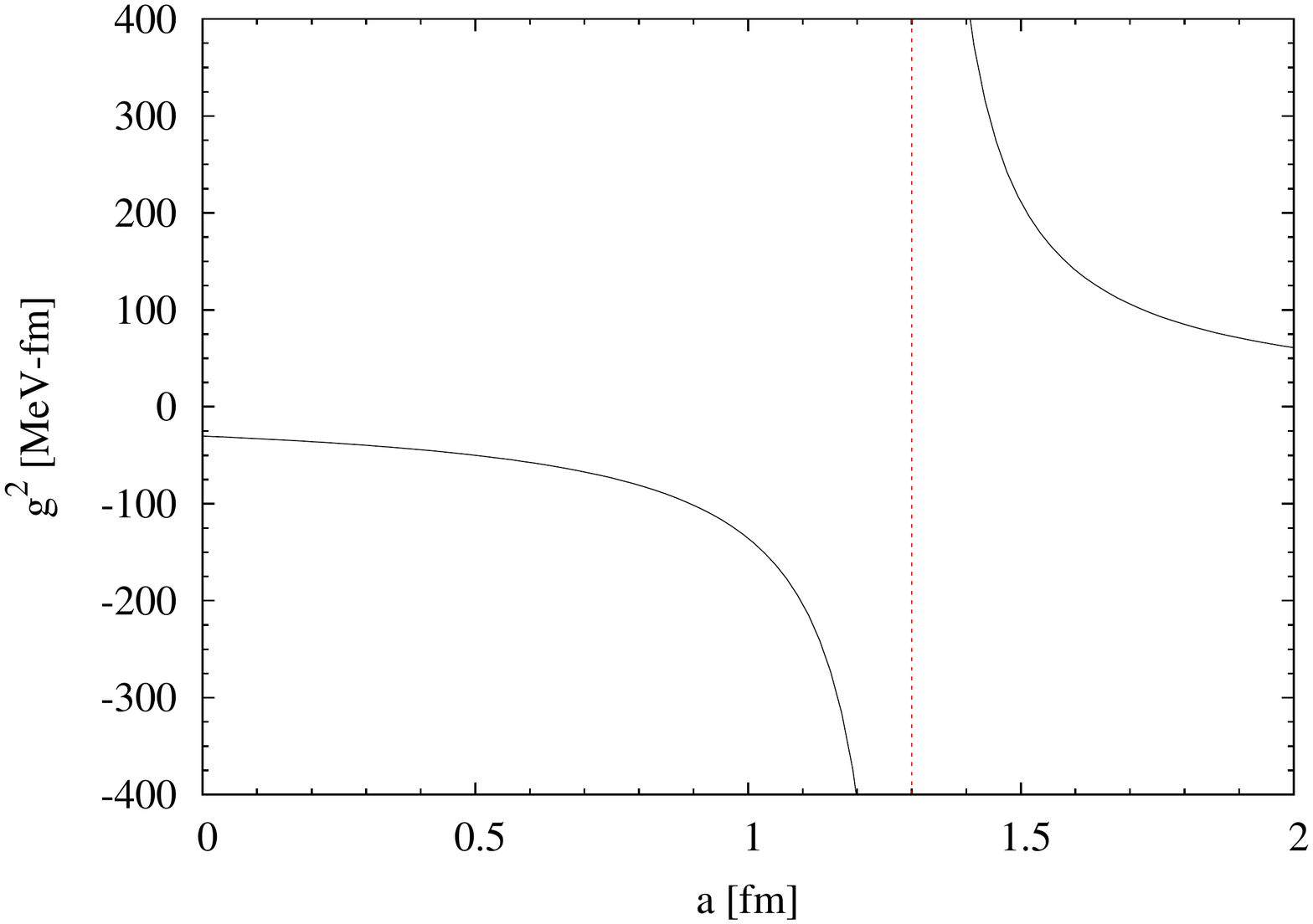}
\caption{\label{fig:npg2vsa}
Behavior of $g^2(a)$ with channel radius $a$, as in
Fig.~\ref{fig:npelvsa}.  Since $- a_0 \simeq 24 \,{\rm fm} \gg 2 \,
{\rm fm}$, the factor $(a - a_0)$ that relates $g^2(a)$ to
$E_\lambda(a)$ [Eq.(\ref{g2})] varies little in the range plotted, and
hence the shape of $g^2(a)$ is nearly the same as that of
$E_\lambda(a)$ shown in the figure above.}  
\end{center} 
\end{figure} 
 
As an explicit demonstration of this behavior we consider the case of
$S$-wave $np$ scattering. In this case the low-energy parameters are
$a_0=-23.7$ fm and $r_0=2.75$ fm \cite{Hackenburg:2006qd}. The
expressions in Eqs.~\eqref{Elambda} and \eqref{g2} are shown in Figs.\ 
\ref{fig:npelvsa} and \ref{fig:npg2vsa}, respectively, for this case.

We see from the figures that both $E_\lambda(a)$ and $g^2(a)$ are
positive above the pole at $a_p=1.30$ fm, and both are negative below
the pole.  At $a=0$, they have the values $E_\lambda(0)=-1.27$ MeV and
$g^2(0)=-30.2$ MeV-fm [$g^2(0)/(\hbar c)=-0.153$].  Since $g^2$ is
negative for $a < a_p$, so is $\gamma^2 = g^2 / a$. Hence, the
conventional $R$-matrix description with $\gamma^2>0$ is not possible
with a channel radius less than $a_p = 1.30$ fm.

A similar situation obtains for the $^{3}\!S_1$ case, where the
scattering length and effective range are 5.411 fm and 1.74 fm
\cite{Hackenburg:2006qd}, respectively. In this case, the pole is
located at 1.07 fm. In distinction to the $^{1}\!S_0$ case, the
scattering length and effective range are both positive. This implies
that the $R$-matrix parameters, $E_\lambda(0)$ and $g^2(0)$, now
take the values $+8.81$ MeV and $-47.7$ MeV-fm [$g^2(0)/(\hbar
c)=-0.242$], respectively.

\section{Two-channel description of the $dt$ reaction} 
\label{sec:dt}
  
The single-level, two-channel formula \cite{LT58} for the $J^\pi=3/2^+$
$^3$H$(d,n)^4$He reaction cross section is
\begin{align}
\label{eqn:1level}
\sigma^{3/2^+}_{n,d} = & \frac{4\pi}{k_d^2}\frac{2}{3}P_nP_d 
\gamma_n^2\gamma_d^2
\left|E_\lambda-E-\gamma_d^2(S_d-B_d)\right. \nonumber \\
      &\left. -\gamma_n^2(S_n-B_n)-i(\gamma_d^2P_d+\gamma_n^2P_n)
\right|^{-2},
\end{align}
with $S_c$ and $P_c$ ($c=n,d$) the real and imaginary parts,
respectively, of the dimensionless outgoing-wave logarithmic derivative, 
\be
\label{eqn:L}
L_c=\frac{a_c}{O_c} \frac{\partial O_c}{\partial r_c}\Bigg|_{r_c=a_c}.
\ee
The boundary condition numbers $B_d$ and $B_n$  are real constants
that are, in principle, arbitrary and can be chosen for convenience,
as discussed below. We will make these choices in order to match the
field-theoretical expression as the radii approach zero, as described
in Section \ref{subsec:zrl}.  For the charged $dt$ channel, the
penetrability, $P_d$ and the shift function, $S_d$ are given in terms
of Coulomb functions by 
\begin{align}
P_d&=\frac{\rho_d}{F_0^2+G_0^2}, \\
S_d&=(F_0F_0^\prime + G_0G_0^\prime) P_d,
\end{align}
where $F_0=F_0(\rho_d,\eta_d)$ and $G_0=G_0(\rho_d,\eta_d)$ are,
respectively, the regular and irregular Coulomb functions for $\ell=0$,
with $\rho_d=k_d a_d$ and $\eta_d=e^2 \mu_d/(\hbar^2 k_d)$.  Here,
$k_d$ is the center-of-mass wave number in the $dt$ channel
(previously called $p_{dt}$ \cite{BH13}), 
$\mu_d=m_dm_t/(m_d+m_t)$ its reduced mass (previously
called $m_{dt}$), and $a_d$ is its channel radius.  The prime means
the derivative with respect to $\rho_d$.  Similar quantities are defined
for the $n\alpha$ channel in terms of the Riccati-Bessel functions
for orbital angular momentum, $\ell=2$, are given as
\begin{align}
P_n &=\frac{\rho_n}{u_2^2+v_2^2}, \\
S_n &=(u_2u_2^\prime+v_2 v_2^\prime)P_n,
\end{align}
with $u_2=\rho_nj_2(\rho_n)$ and $v_2=-\rho_nn_2(\rho_n)$, $j_2$ and
$n_2$ being the ordinary regular and irregular spherical Bessel
functions.  Here the prime means derivative with respect to
$\rho_n=k_n a_n$, with $k_n$  the center-of-mass
wave number (previously
$p_{n\alpha}$) and channel radius $a_n$ in the $n\alpha$ channel.

Aside from the channel radii $a_c$, the remaining $R$-matrix
parameters in the single-level expression \eqref{eqn:1level} are
the reduced-width amplitudes $\gamma_c$ for $c=d,n$, and the energy
eigenvalue $E_\lambda$. All of these are real parameters as a result
of the assumed Hermiticity of the interaction Hamiltonian and the
chosen conditions on the wave functions at the boundary of the
interior region.

\subsection{Finite channel radii}
\label{subsec:fcr}

The original formulation of $R$-matrix theory of Wigner and
Eisenbud \cite{WE47} is based upon the separation into interior
(strongly interacting) and exterior (non-strong) regions of the
configuration space of nucleons for each channel partition (pair).
$R$-matrix theory is, by this formulation, a finite-range ($a_c > 0$)
description of nuclear reactions.  However, as we will see in the
following section, one may sensibly take the zero-radius
limit of the theory. In so doing, we reproduce 
the EFT expressions for the cross section [see Eq.\eqref{eqn:signd}]
that follow from a quantum effective field theory of particles 
interacting via local interactions.  As shown in the companion
paper \cite{BH13}, a high-quality fit to the $^3$H$(d,n)^4$He reaction
data within the EFT treatment required the sign of the energy shift,
$\Re [\Sigma_{dt}^{(C)}(W)]$, to be changed.  This led to the use of
the ``wrong-sign'' Lagrangian in the EFT approach, as described there.
The wrong-sign Lagrangian can be alternatively and equivalently
interpreted as having pure-imaginary coupling constants.  Based on the
identity of the EFT and $R$-matrix expressions for the $^3$H$(d,n)^4$He
reaction cross section (see below), we anticipated that, as the
$R$-matrix radii are taken to zero, the reduced-width amplitudes
$\gamma_c$ would become pure imaginary numbers. It is instructive to
study the nature of the transition from real to pure-imaginary values
of the $\gamma_c$ in $R$-matrix theory, where the full two-channel $R$
matrix is used, in order to better understand its relation to an
``unphysical" Lagrangian in EFT.

We therefore conducted a numerical study of the dependence on channel
radii of the two-channel $R$-matrix fit to the $dt$ reaction cross
section.  The cross section data fitted with Eq.\eqref{eqn:1level}
were the same ones \cite{Arn54,Jar84,Brn87} used for the EFT fit, 
employing a {\em Mathematica} program that could easily be extended to 
complex values of some of the parameters.  It was found that no meaningful
reduction in the $\chi^2$ was achieved by allowing separate values of
the channel radii, so the fits were made with $a_d=a_n=a$.   The boundary condition numbers $B_c$ were taken to be the energy-independent part of the shift function, as given by Eq.\eqref{eqn:Slinfa} in the Appendix.  This gives $B_n=-2$ independent of $a$, but $B_d=- x_0K_0(x_0)/(2K_1(x_0))$,  $K_n$ being the irregular modified Bessel function evaluated at $x_0=\sqrt{8a/b_0}$, which depends on $a$.  Here, $b_0=34.62$ fm is a length for the $dt$ system equivalent to the Bohr radius.

\begin{figure}[htbp]
\begin{center}
\includegraphics[width=3.25in]{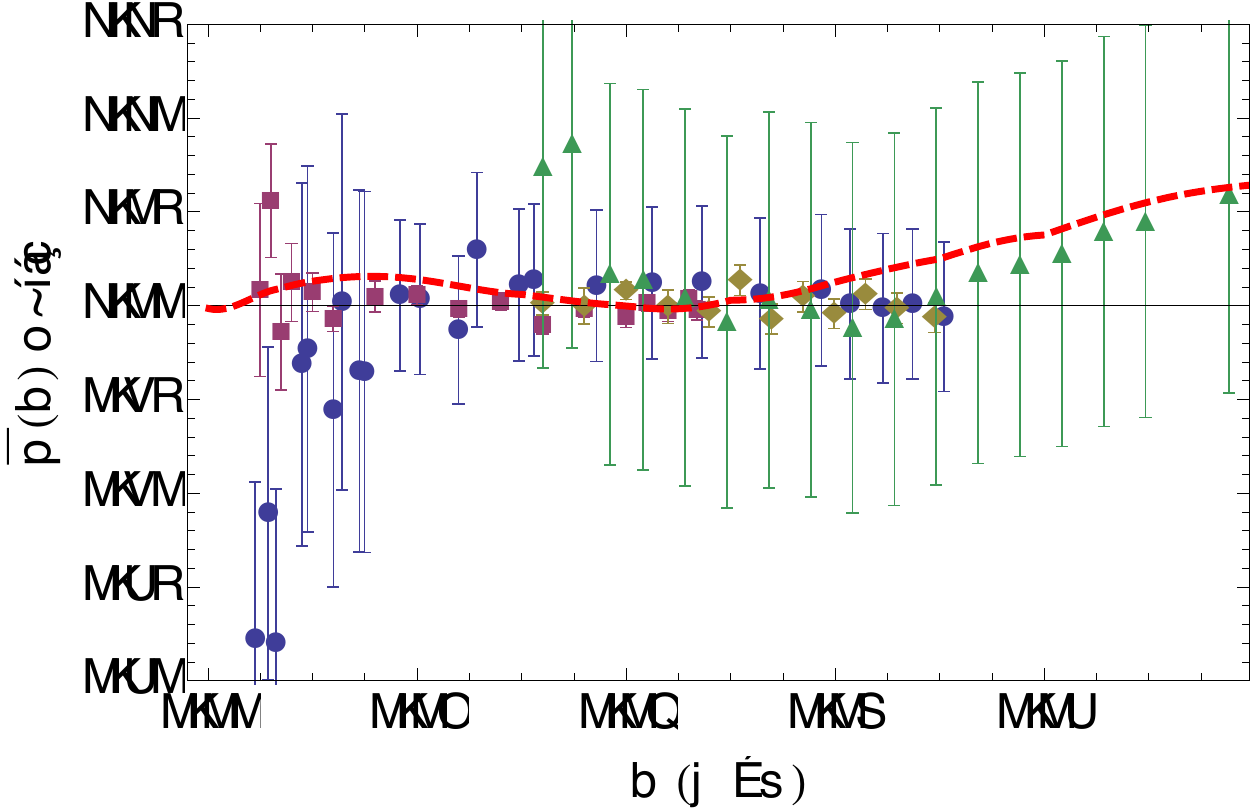}
\caption{The $d+t$ reaction data divided by the single-level,
four-parameter $R$-matrix fit described in the text. The (blue) circles are
the data of Arnold {\em et al.}~\cite{Arn54}; the (magenta) squares
are the data of Jarmie {\em et al.}~\cite{Jar84} renormalized by a
factor of 1.017; the (olive) diamonds are the relative data of Brown
{\em et al.}~\cite{Brn87} renormalized by a factor of 1.025; the
(green) triangles are the data of Argo {\em et al.}~\cite{Arg52},
which were not included in the single-level fit because of their large
error bars.  This fit yields a
$\chi^2$ per degree of freedom of 0.713 which is to be compared with 
the value  of 0.784 determined in the three-parameter EFT fit of 
Ref.\cite{BH13}.  The red dashed curve presents the results of the
$R$-matrix fit of Bosch and Hale~\cite{B&H92} divided by the single-level
four parameter fit. }
\label{fig:ratio}
\end{center}
\end{figure}

The best fit ($\chi^2= 34.94$ corresponding to a $\chi^2/$DOF = 0.713)
was obtained for $a=7$ fm, although $\chi^2$ was a shallow function of
$a$ in the range $a=3$ to 8 fm.  The best-fit parameters for $a=7$ fm
and boundary conditions $B_d=-0.59, B_n=-2$ are: $E_\lambda= 179 \pm
5$ keV, $\gamma_d^2=324 \pm 12$  keV, $\gamma_n^2=12.2 \pm 0.2$ keV.
This best single-level fit to the experimental $dt$ data is shown in
Fig.~\ref{fig:ratio}, which displays the data divided by the
theoretical fit.  Thus the theory appears simply as the horizontal
line at the ordinate 1.00.  (All the quantities were first expressed
in terms of the dimensionless astrophysical $S$ factor $\overline{S}$
defined in Eq.(1.11) of Ref.\cite{BH13}.)  Also shown are the results
of Bosch and Hale (BH) \cite{B&H92}  divided by the single-level fit.
They are close to the single-level fit at low energies, but their
ratio to it increases at higher energies.

For solutions near $a=2$ fm, the three fitting parameters became
quite large in magnitude, while maintaining the same signs they had at 
larger radii.  However, for $a < 2$ fm, roughly comparable fits were obtained 
to the data with the signs of all the parameters changed ($E_\lambda < 0, 
\gamma_c^2 < 0$).  This behavior is consistent with having gone through a 
pole at $a_p$ somewhat less than 2 fm, and having pure-imaginary 
reduced-width amplitudes $\gamma_c$ at zero radius, in qualitative 
agreement with the wrong-sign Lagrangian EFT result, and similar to the 
case of $np$ scattering discussed in Sec.\ref{sec:np}. In the next section, we will show that this agreement between the $R$-matrix description at zero
radius and the EFT result is exact.

\subsection{Taking the zero-radius limit}
\label{subsec:zrl}
In order to match to the EFT expression, including a single 
unstable, intermediate field, for the cross section Eq.~(4.5)
from Ref.\cite{BH13}, we make the associations
\be
\gamma_d^2= -\frac{g_d^2}{2\pi}\frac{\mu_d}{\hbar^2 a_d}~{\rm and}~
\gamma_n^2= -\frac{g_n^2}{6\pi}\frac{\mu_n}{\hbar^2 a_n^5}
\ee
between the reduced widths $\gamma_c^2$ and squared EFT coupling
constants $g_c^2$, and let the channel radii $a_c$ approach
zero.\footnote{The small $a_c$ behavior of the reduced widths can be
  understood as follows: at small values of $a_c$, the radial wave
  function is dominated by the irregular solution, so that $u_l(a_c)\sim
  a_c^{-l}$, and $\gamma_{cl}^2=\frac{\hbar^2}{2\mu a_c}u_l^2(a_c)\sim
  1/a_c^{2l+1}$. \label{foot}} 
\label{smallac} 
The minus signs are necessary to account for the
wrong-sign Lagrangian convention used in Ref.\cite{BH13}, in
which the coupling constants $g_c$ are assumed to be real, rather than
pure imaginary.  Although in this small-$a_c$ 
limit the $R$-matrix reduced-width
amplitudes $\gamma_c$ become infinite as the penetrabilities $P_c$
approach zero, we expect the ``half-width" terms in both the numerator
and denominator of the cross section expression
\begin{align}
   \gamma_d^2 P_d &\rightarrow 
-\frac{g_d^2}{2\pi}\frac{\mu_d}{\hbar^2}k_dC_0^2(\eta_d),\\
\gamma_n^2P_n &\rightarrow 
-\frac{g_n^2}{6\pi}\frac{\mu_n}{\hbar^2}k_n^5,
\end{align}
to remain finite.  Here, $C_0^2(\eta)=2\pi\eta[\exp(2\pi\eta)-1]^{-1}$
is called $|\psi_{{\bf p}_{dt}}^{(C)}(0)|^2$ in the companion paper.
Additionally, we choose the channel-surface boundary conditions,
$B_{n,d}$, to be the energy-independent part of the shift functions at
zero energy, $S_c^\infty = S_{\ell(c)}(\infty,a)$, as is written in the
Appendix.  That is, we choose $B_n= S_n^\infty$ and $B_d=S_d^\infty$,
so that the energy shift of $E_\lambda$ in the denominator of
Eq.~\eqref{eqn:1level},
\begin{align}
   \Delta &=-\gamma_d^2(S_d-B_d)-\gamma_n^2(S_n-B_n),
\end{align}
depends on quantities $\tilde{S_c}=S_c-S_c^\infty$ that satisfy a
dispersion relation (see the Appendix).  $\tilde{S_n}$ vanishes in the
$n\alpha$ channel, leaving only the energy-dependent shift in the $dt$
channel, given according to Eqs.~\eqref{eqn:ReLd} of the Appendix and
Section \ref{subsec:fcr} above by
\be
\Delta_d(E)=-\gamma_d^2\tilde{S_d}\rightarrow  
\frac{g_d^2}{\pi}\frac{\mu_d}{\hbar^2 b_0}[\Re\psi(i\eta_d)-\ln(\eta_d)].
\ee 
 We note that the associations above allow us to connect with
the Coulomb self-energy of the previous paper\cite{BH13},
\be
\gamma_d^2\bar{L}_d \rightarrow -\frac{g_d^2}{2\pi}\frac{\mu_d}{\hbar^2 
a_d}
(\bar{S}_d+iP_d)=\Sigma_{dt}^{(C)}.
\ee

Using these limiting values, the $R$-matrix cross section expression,
Eq.\eqref{eqn:1level} becomes
\begin{align}
\label{eqn:signd}
\sigma^{3/2^+}_{n,d}
&=\frac{32\pi}{9\hbar v_d}\frac{g_d^2}{4\pi}\frac{g_n^2}{4\pi}
\frac{\mu_n}{\hbar^2}k_n^5 C_0^2(\eta_d) \nonumber \\
&\times\Bigg|E-E_\lambda-\Delta_d(E) \nonumber \\
&-i\left[\frac{g_d^2}{2\pi}\frac{\mu_d}{\hbar^2}k_dC_0^2(\eta_d)
+\frac{g_n^2}{6\pi}\frac{\mu_n}{\hbar^2}k_n^5\right]\Bigg|^{-2},
\end{align}
which is in complete agreement with Eqs.~(1.6)--(1.9) in
Ref.\cite{BH13}, keeping in mind that $E=\frac{{\bf
p}_{dt}^2}{2m_{dt}}$, $E_\lambda=E_*$, and
$\Delta_d(E)=\frac{g_{dt}^2}{4\pi}\Delta(W)$.

\section{discussion and conclusions}
\label{sec:dc}

This investigation has revealed some interesting, and possibly
significant, connections of single-level $R$-matrix theory to other
theoretical approaches.  It is apparent that the experimental data for
the two cases considered, $np$ scattering and the $dt$ reaction,
dictate minimum channel radii at which the $R$-matrix parameters are
physical.  For $np$ scattering, this radius is just above 1 fm, and
for the $dt$ reaction, it is just below 2 fm.  This finding is
consistent with the presumptive connection of the channel radii with
the range of nuclear forces.  What is surprising, however, is that one
can continue the $R$-matrix parameters below the pole that occurs at
these minimum radii [{\em c.f.} Figs.~\ref{fig:npelvsa} and 
\ref{fig:npg2vsa}], and obtain equally good fits to the experimental
data with negative reduced widths $\gamma^2$. The continuation to
zero radius done in this way gives identically the same result as does
effective field theory with a wrong-sign Lagrangian, in which 
local interactions are mediated by a single unstable field. 

While it appears that the position of the poles is related to the
range of nuclear forces, we currently lack a complete understanding of
the physical significance of the pole in the dependence of the
single-level $R$-matrix parameters on channel radius. However, the
presence of this singularity separating the physical and non-physical
descriptions of the experimental data using $R$-matrix theory gives an
indication of how to interpret the wrong-sign Lagrangian in effective
field theory: imaginary coupling constants in the field theory, which
are equivalent to the wrong-sign free-field Lagrangian, appear to
compensate for a description of the finite-range (nuclear) forces with
zero-range interactions. A direct correspondence of the
reduced-width amplitudes in $R$-matrix theory to the coupling
constants of EFT as discussed in Section \ref{sec:dt} of this paper
appears in the limit of vanishing channel radii.

For the $dt$ reaction, the extrapolation to zero radius gives a
description that is similar to the ``model-independent" effective
range expansion of Karnakov {\sl et al.} \cite{Karnakov90}.  However,
the description given here and in the companion paper \cite{BH13}
requires only three parameters, whereas that of Ref.\cite{Karnakov90}
employs four parameters, which is the number required for a
single-level $R$-matrix description at finite radii.

The present study establishes an identity between the EFT treatment of
light nuclear reactions with an unstable intermediate field using
local interactions with that of the single-level, two-channel $R$
matrix in the limit in which the channel radii are taken to zero. We
have found that poles in the level energy and channel widths appear in
the continuation in channel radii between this limit and radii that
correspond to real, physical values of the widths.  The question of
the physical interpretation of such poles is beyond the scope of this
work. Among the questions raised by the poles is the issue of whether
the poles arise due to the restriction to a single-level $R$-matrix
description. We are currently studying this question using potential
models.

\acknowledgments{This work was carried out under the auspices of the
National Nuclear Security Administration.}

\appendix*

\section{Dispersion relations for the outgoing-wave logarithmic derivative}

A quantity of central importance in this discussion is the
outgoing-wave logarithmic derivative of Eq.~\eqref{eqn:L},
\be
\label{eqn:Lc}
L_c=\frac{a_c}{O_c} 
\frac{\partial O_c}{\partial r_c}\Bigg|_{r_c=a_c}=S_c+iP_c.
\ee
For charged-particle channels such as  $dt$, the outgoing-wave
solution is defined in terms of Coulomb functions by
\be
O_d=(G_0+iF_0)\exp(-i\sigma_0),
\ee
with $\sigma_0=\arg\Gamma(1+i\eta_d)$ the $S$-wave Coulomb phase
shift.   For the neutral-particle $n\alpha$ channel, it is defined 
in terms of the Riccati-Bessel functions for $\ell=2$,
\be
O_n=-i\rho_n[j_2(\rho_n)+in_2(\rho_n)] = -i\rho_n h^+_2(\rho_n),
\ee
where
$j_2$ and $n_2$ are the ordinary regular and irregular spherical
Bessel functions, respectively, $h^+_2$ is the outgoing Hankel function
of second order, and $\rho_n=k_n a_n$, the product of the wave
number in the center-of-mass, $k_n$ and channel radius $a_n$ in 
the $n\alpha$ channel.

It is useful for this application to develop a dispersion relation for
the real and imaginary parts of $L_c$.  Ordinarily, this would be done
by means of a Hilbert transform that depends on the analytic
properties of $L_c(E)$ in the cut $E$-plane. The fact that the the
outgoing-wave logarithmic derivative diverges as
$|E|\rightarrow\infty$ [see Eq.\eqref{eqn:L2rho} below], however,
prevents a direct application of the Hilbert transform in the
complex-$E$ plane. Since this function is finite for all values of
$\ell$ for $E=0$, it makes sense to consider a Hilbert transform for
it in the {\sl inverse} energy. This procedure is slightly different
for Hankel functions (neutral-particle channels) and Coulomb functions
(charged-particle channels), so they will be addressed separately.

We first consider the neutral case and explicitly display only the
dependence of the channel $c$ [as in Eq.\eqref{eqn:Lc}]
on the orbital angular momentum $\ell$.
The $n\alpha$ $\tfrac{3}{2}^+$ channel, for example, has $\ell=2$, and
we examine $L_2(\rho_n)=\rho_nh_2^{+'}(\rho_n)/h_2^+(\rho_n)$, where
the prime means differentiation with respect to $\rho_n$.  It can be
written as
\be
\label{eqn:L2rho}
L_2(\rho_n)=-2+\frac{1-i\rho_n}{\frac{3}{\rho_n^2}-\frac{3i}{\rho_n}-1},
\ee
which shows that  $\tilde{L}_2(\rho_n)=L_2(\rho_n)+2$ is finite 
(zero, in fact) at infinity in $x=\rho_n^{-2}$.  
A less obvious consequence of using
inverse energy variables is that, whereas the integration contour for
the Hilbert transform would be taken on the first sheet in energy
variables, it must be taken on the second sheet in the inverse-energy  
variable $\rho_n^{-2}$.  Furthermore, since on that sheet the positive
$\rho_n^{-1}$ axis lies on the bottom rim of the cut, we must approach
the cut {\sl from below} to get the physical values. Therefore, the
Hilbert transform for $\tilde{L}_2(\rho_n)$ has the form in the cut
$x$-plane of
\be
\label{eqn:L2trho}
 \tilde{L}_2(\rho_n)=-\frac{1}{\pi}\int_0^\infty dx 
 \frac{\Im\tilde{L}_2(x)}{x-\rho_n^{-2}+i\epsilon}.
\ee
Using the familiar Plemelj relation
\be
\frac{1}{x-x_0+ i\epsilon}=\frac{P}{x-x_0}- i\pi\delta(x-x_0) \,,
\ee 
the above expression reduces to an identity for the imaginary part
of $\tilde{L}_2$, but for the real part yields the relation:
\begin{align}
\Re\tilde{L}_2(\rho_n)&=S_2(\rho_n)+2 \nonumber \\
&=-\frac{1}{\pi}\dashint dx \frac{P_2(x)}{x-\rho_n^{-2}},
\end{align}
where $\dashint$ is the Cauchy principal value.
This result can be generalized for uncharged channels to the desired
dispersion relation between the real and imaginary parts of $L_\ell$,
\be
\label{eqn:Slrho}
S_\ell(\rho)=-\ell-\frac{1}{\pi}\dashint dx \frac{P_\ell(x)}{x-\rho^{-2}}.
\ee
As a check for $\ell=2$, we write Eq.~\eqref{eqn:L2rho} in terms of
$x$ and resolve it into real and imaginary parts:
\be
\tilde{L}_2(x)=\frac{3x+2+ix^{-1/2}}{9x^2+3x+1}.
\ee
Then Eq.~\eqref{eqn:L2trho}
implies the relation between the real and imaginary parts
\begin{align}
&\frac{3y+2}{9y^2+3y+1} \nonumber \\
&=-\frac{1}{\pi}\dashint  dx\frac{x^{-1/2}}{(x-y)(9x^2+3x+1)},
\end{align}
which is easily verified by performing the principal-value integral. 

For Coulomb functions, the natural inverse-energy variable to use for
the Hilbert transform is $\eta^2=\frac{e^2}{2b_0E}$, with
$b_0=\frac{\hbar^2}{e^2\mu}$ the equivalent of the Bohr radius for the
initial pair of charged particles.  Then the conventional variable
$\rho=ka$ is determined by $k=(\eta b_0)^{-1}$.  Therefore, we expect
the Hilbert transform and a dispersion relation analogous to
Eqs.~\eqref{eqn:L2trho} and \eqref{eqn:Slrho} for Coulomb functions at
a finite radius $r=a$ to be
\begin{align}
\label{eqn:Lletaa}
L_\ell(\eta,a)&=S_\ell(\infty,a)\nonumber \\
&-\frac{1}{\pi}
\int_0^\infty dx\frac{P_\ell(x,a)}{x-\eta^2+i\epsilon},\\
\label{eqn:Sletaa}
S_\ell(\eta,a)&=S_\ell(\infty,a)
               -\frac{1}{\pi}\dashint dx\frac{P_\ell(x,a)}{x-\eta^2},
\end{align}
where $S_\ell(\infty,a)$ is a real constant that gives the value of
$L_\ell$ at infinite $\eta$ (zero energy).  For Hankel functions, this
is simply $S_\ell(\infty)=-\ell$, but for Coulomb functions it is
given by 
\be
\label{eqn:Slinfa}
S_\ell(\infty,a)=-\ell-\frac{x_0}{2}\frac{K_{2\ell}(x_0)}{K_{2\ell+1}(x_0)},
\ee 
with $K_n(x_0)$ the irregular modified Bessel function evaluated
at $x_0=\sqrt{8a/b_0}$.

The validity of these relations is difficult to test in general, but
one of the results in our companion paper [Eq.(4.19)]
is a special case of this
Coulomb function dispersion relation.  In that case, we want to find
the shift function that belongs with the $dt $ penetrability function
at vanishingly small radius $a_d$,
\be
\label{eqn:Psma}
P_d(\eta) \approx \rho_dC_0^2(\eta)=\frac{2\pi a_d/b_0}{\exp(2\pi\eta)-1}.
\ee 
This sort of penetrability function occurs in the integral
representation of the digamma function,
\begin{align}
\psi(z)&=\ln(z)-\frac{1}{2z} \nonumber \\
&-\int_0^\infty dt^2\frac{1}{t^2+z^2}\frac{1}{\exp(2\pi t)-1}.
\end{align}
Letting $z= i\eta+\epsilon/(2\eta)$, with $\epsilon$ a positive
infinitesimal ($\Re z>0$ is
required for the validity of the above expression), we have
\begin{align}
\label{eqn:psiieta}
\psi(i\eta)&=\ln(i\eta)-\frac{1}{2i\eta}\nonumber\\
&-\int_0^\infty dt^2\frac{1}{t^2-\eta^2+i\epsilon}\frac{1}{\exp(2\pi t)-1},
\end{align}
which has the form of Eq.~\eqref{eqn:Lletaa} for the penetrability in 
Eq.~\eqref{eqn:Psma}, with $x=t^2$ and $l=0$. 

The shifted logarithmic derivative,
\begin{align}
\tilde{L}_d=S_d-S_d^\infty+iP_d,
\end{align}
with
\begin{align}
S_d^\infty &= S_0(\eta\rightarrow\infty,a_d\rightarrow 0)\nonumber\\
&=(2a_d/b_0)[\ln(2a_d/b_0)+2\gamma]\rightarrow 0,
\end{align}
can therefore be expressed as the Hilbert transform based on rearranging 
Eq.~\eqref{eqn:psiieta},
\begin{align}
\frac{b_0}{2a_d}\tilde{L}_d(\eta) 
&=-\int_0^\infty dt^2\frac{1}{t^2-\eta^2+i\epsilon}\frac{1}{\exp(2\pi t)-1}
\nonumber \\
&=\psi(i\eta)-\ln(i\eta)+\frac{1}{2i\eta}.
\end{align}
Using properties of the digamma function \cite{AS}, we have 
\be
\frac{b_0}{2a_d}\Im\tilde{L}_d(\eta)
=\Im\psi(i\eta)-\frac{\pi}{2}-\frac{1}{2\eta},
\ee
and since
\be
\Im\psi(i\eta)
=\frac{1}{2\eta}+\frac{\pi}{2}\frac{\exp(2\pi\eta)+1}{\exp(2\pi\eta)-1},
\ee
it gives the check that
\begin{align}
\frac{b_0}{2a_d}\Im\tilde{L}_d(\eta)
&=\frac{\pi}{2}\left(\frac{\exp(2\pi\eta)+1}{\exp(2\pi\eta)-1}-1\right)
\nonumber \\
&=\frac{\pi}{\exp(2\pi\eta)-1}=\frac{b_0}{2a_d}P_d.
\end{align}
The new relation we want, however, is given by
\begin{align}
\label{eqn:ReLd}
\frac{b_0}{2a_d}\Re\tilde{L}_d(\eta)
&=\frac{b_0}{2a_d}\tilde{S_d}\nonumber \\
&=\Re\psi(i\eta)-\ln(\eta)=h(\eta),
\end{align}
which agrees with the energy-dependent part of direct expansions ({\sl
e.g.}, Jackson and Blatt \cite{JB}) of the $\ell=0$ Coulomb wave
functions for small $a_d$, and with Eq.~(4.19) of Ref.\cite{BH13}.


\end{document}